\documentclass[sigconf]{acmart}

\usepackage[noend]{algorithmic}
\usepackage{algorithm}
\usepackage{comment}
\usepackage{graphicx}
\usepackage{subfig}
\usepackage{ltablex}
\usepackage{amsmath}

\def\BibTeX{{\rm B\kern-.05em{\sc i\kern-.025em b}\kern-.08emT\kern-.1667em\lower.7ex\hbox{E}\kern-.125emX}}

\copyrightyear{2019} 
\acmYear{2019} 
\setcopyright{rightsretained} 
\acmConference[arXiv]{arXiv}
\acmBooktitle{2019}

\begin{document}

\title[ROMEO]{ROMEO: A Plug-and-play Software Platform of Robotics-inspired Algorithms for Modeling Biomolecular Structures and Motions}

\author{Kevin Molloy}
\affiliation{%
  \institution{Dept of Computer Science}
  \institution{George Mason University}
  \streetaddress{Fairfax, VA 22030}}
\author{Erion Plaku}
  \affiliation{%
  \institution{Department of Computer Science}
  \institution{The Catholic University of America} 
  \streetaddress{Washington, D.C., 200064}}
\author{Amarda Shehu}
\authornote{Corresponding Author}
\affiliation{%
  \institution{Dept of Computer Science}
  \institution{George Mason University}
  \streetaddress{Fairfax, VA 22030}}
\email{amarda@gmu.edu}
  
\renewcommand{\shortauthors}{K. Mollloy, E. Plaku, and A. Shehu}

\begin{abstract}
  
\textbf{Motivation:} Due to the central role of protein structure in
molecular recognition, great computational efforts are devoted to
modeling protein structures and motions that mediate structural
rearrangements. The size, dimensionality, and non-linearity of the
protein structure space present outstanding challenges. Such
challenges also arise in robot motion planning, and robotics-inspired
treatments of protein structure and motion are increasingly showing
high exploration capability. Encouraged by such findings, we debut
here ROMEO, which stands for Robotics prOtein Motion ExplOration
framework. ROMEO is an open-source, object-oriented platform that
allows researchers access to and reproducibility of published
robotics-inspired algorithms for modeling protein structures and
motions, as well as facilitates novel algorithmic design via its
plug-and-play architecture. \\ \textbf{Availability and
  implementation:} ROMEO is written in C++ and is available in GitLab
(https://github.com/).  This software is freely available under the
Creative Commons license (Attribution and
Non-Commercial). \\ \textbf{Contact:}
\href{amarda@gmu.edu}{amarda@gmu.edu}\\

\end{abstract}

\maketitle

\section{Introduction}

Protein molecules regulate virtually all processes that maintain and
replicate a living cell, and their tertiary structure largely
determines their interactions with molecular
partners~\cite{BoehrWright08}. A detailed understanding of the
structure(s) at the disposal of a protein for biological activity and
of the motions that mediate rearrangements between different
structures for activity modulation is central to understanding
molecular mechanisms~\cite{BoehrWright09}. Decades of research have
shown that such understanding cannot be obtained via experimentation
or computation alone; in particular, while great progress is made on
modeling biomolecular structures and motions, challenges related to
the size, dimensionality, and non-linearity of structure spaces
associated with macromolecules, such as proteins,
remain~\cite{MaximovaShehuPCB16}. Over the past decade, great
advancements in the ability to explore complex protein structure
spaces have come from robotics-inspired algorithms that leverage
analogies between molecular and robot configuration
spaces~\cite{ShehuPlakuJAIR16}.

Encouraged by such progress, we debut here ROMEO, which stands for
Robotics prOtein Motion ExplOration framework. ROMEO is an
open-source, object-oriented platform that allows researchers access
to and reproducibility of published robotics-inspired algorithms for
modeling protein structures and motions. For instance, ROMEO provides
templates for popular robotics motion planning algorithms, such as the
Rapidly-exploring Random Tree (RRT) and Probabilistic RoadMap (PRM),
and offers disseminated adaptations of these canonical algorithms that
address diverse application settings, from template-free protein
structure prediction to modeling of motions that mediate
re-arrangements between stable and meta-stable
structures~\cite{AmatoDill02, ShehuOlsonIJRR10,
  MolloyShehuBMCStructBiol13, devaurs_characterizing_2015}.


\section{Plug and Play Design}

ROMEO follows a plug-and-play architecture to facilitate novel
algorithmic design and so allow researchers to further advance
algorithmic research in molecular biology. ROMEO is written in C++ and
its object-oriented design allows easy adaptation and expansion of its
classes. These classes have been designed around a core set of
components shared by many motion planning algorithms, which we
summarize below (see Figure S1 for a visualization of the
architecture).

{\bf Representation, Energy, and Forward Kinematics:} The {\it cfg}
class represents a protein configuration.  Included with this initial
release is an extension of this class that interfaces with the Rosetta
package \cite{Bradley05} and utilizes the coarse-grained
representation known as the centroid mode, which tracks heavy backbone
atoms and a side-chain centroid pseudo-atom via dihedral
angles. Developers can easily extend this class to support Rosetta's
all-atom representation or others. A forward kinematics class allows
projecting a configuration (under the selected representation) into
Cartesian space space (retrieving Cartesian coordinates for each of
the represented atoms). This class also associates an energy/fitness
score with a configuration.  In this initial release, ROMEO utilizes
Rosetta scoring functions.

{\bf Planners:} Sampling-based algorithms are popular in robot motion
planning due to their ability to handle high-dimensional configuration
spaces.  ROMEO provides base classes for tree- and roadmap-based
planners. This initial release includes support for RRT, PRM, and
variants (the reader is referred to the review
in~\cite{ShehuPlakuJAIR16} for a description of these
algorithms). These planners share a common set of operations, some of
which we highlight below. (i) ROMEO provides {\it samplers} to create
new configurations and {\it offspring generators} to extend the tree
or roadmap in a target direction.  Producing new samples or offspring
using traditional robotic techniques results in high-energy,
unrealistic configurations. ROMEO offers {\it offspring generators}
that employ molecular fragment replacement, which remedies this
issue~\cite{Bradley05}. (ii) Acceptors are responsible for
determining whether configurations are to be added to the
roadmap/tree.  ROMEO provides examples of extending this class via an
energy threshold or dynamic techniques utilizing the Metropolis. (iii)
ROMEO provides distance classes to measure the distance between two
configurations, supporting both angle-based and Cartesian
coordinate-based representations. (iv) Edge cost evaluators are
responsible for weighting each of the transitions encoded into the
graph/tree.  ROMEO provides many extensions to a baseline class to
support encoding basic distances, energetic differences, and those
based on mechanical work~\cite{devaurs_characterizing_2015}. (v) A
goal acceptor class determines if a given configuration is similar
enough to a given goal configuration and has connectivity to a given
start configuration.

\section{Usage examples}

We showcase two selected examples of how ROMEO can be utilized for
modeling structures and motions.

{\bf Structure Prediction:} We showcase the plug-and-play nature of
ROMEO by the FeLTR algorithm proposed in~\cite{Shehu_IJRR_2010} for
decoy generation in template-free protein structure prediction. FeLTR
grow a search tree in the protein's configuration space to search for
low-energy configurations. To implement FeLTR, the {\it planner} class
is extended to utilize a low-dimensional projection to guide the
exploration. This projection has been shown to strike a good balance
between breadth (the unexplored frontier) and depth (low-energy). Only
300 lines of code were required to extend ROMEO to implement FeLTR
(source code and examples of running this method are included in the
software distribution). Section S1.1 in the Supplementary Information
showcases results.

{\bf Motion Computation:} Tree-based planners have been used to
compute energetically-feasible paths connecting known structures of a
protein.  ROMEO includes an example of such an algorithm known as
SPRINT \cite{MolloyShehuBMCStructBiol13}. SPRNT grows a tree in the
configuration space from a start configuration and biases its
selection based on a low-dimensional projection based on a progress
coordinate to the goal configuration. The software distribution
includes an example of utilizing SPRINT on the cyanovirin-n protein,
where the goal is to compute motions that connect two given structures
more than $16$\AA apart from each-other. Details and results are
available in Section S1.2 of the Supplementary Information.

\section{Applicability}

The examples above illustrate the power of ROMEO's plug-and-play
architecture. This software can be useful in advancing algorithmic
research structure and motion computation.  It can also be employed in
a classroom setting to allow instructors to initiate students in
computational molecular biology and easily extend components of the
framework.

\section{Acknowledgements}
This work has been partially supported by the National Science
Foundation Grant No.1440581.

\section*{APPENDIX}

\section*{ROMEO Design/Classes}

ROMEO is written in C++ and consists of a core set of components that
are shared among all sampling-based robot motion planning
algorithms. The object-oriented design of ROMEO allows easy adaptation
and expansion of its core components, making it possible for
developers to customize ROMEO for additional applications.

\begin{figure*}
\centering
\includegraphics[width=\linewidth]{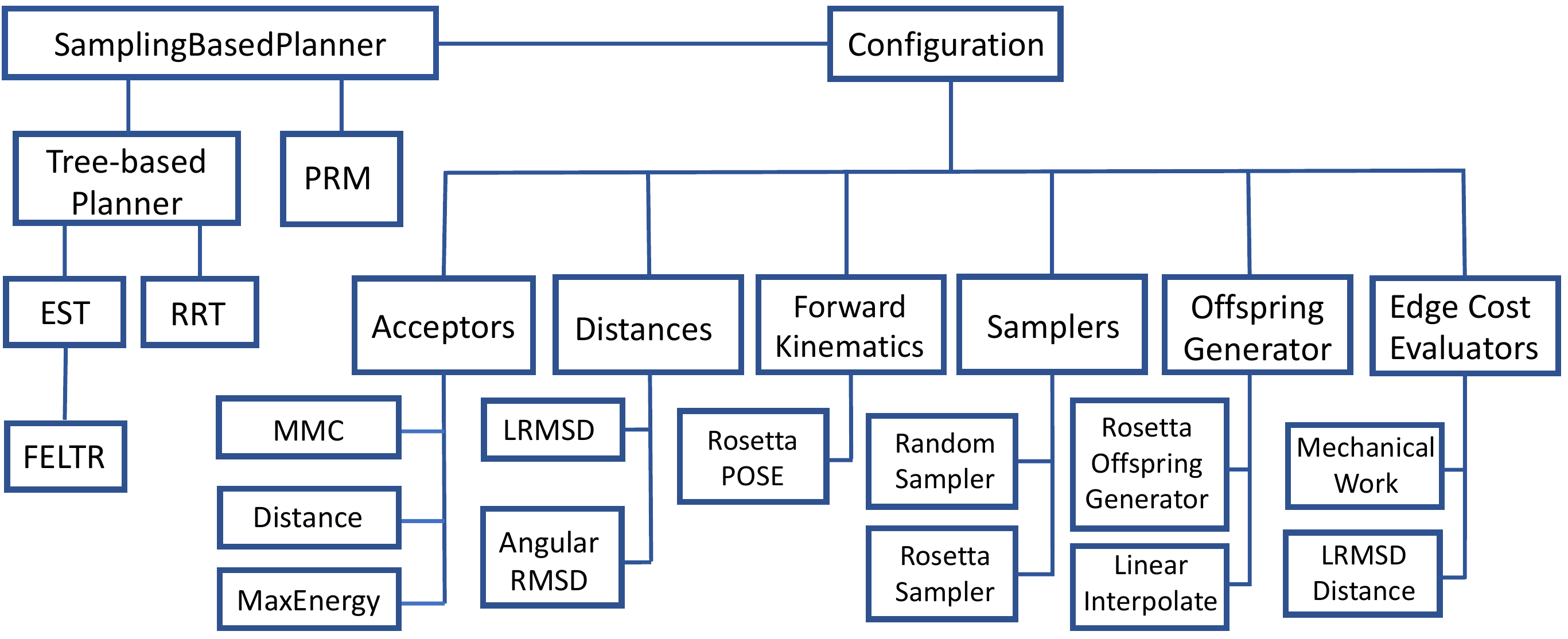}
\caption{In ROMEO's plug-and-play design, all classes are derived from
  a common Configuration object (which can also be overloaded. All of
  the planner functions can also be replaced with new, derived
  classes.}
\label{fig:ROMEO_DESIGN}
\end{figure*}

Figure~\ref{fig:ROMEO_DESIGN} illustrates the class inheritance
hierarchy of the software. The following sections summarize the
purpose of each class and describe in greater detail selected member
functions.

\subsection*{Choice of Representation}

The {\it cfg} class is the base class to represent a
configuration. The class stores a vector of the backbone dihedral
angles of a protein structure and in doing so assumes an idealized
geometry (where bond lengths and valence angles do not deviate from
equilibrium values). This choice of representation is popular in
template-free protein structure prediction and motion computation. We
point that other applications, such as folding, necessitate extending
the base class by allowing bond lengths and valence angles to deviate
(thus including them in the representation) or by introducing other
configuration parameters.

Since the potential energy associated with a given protein structure
is used in many planning operations (for instance, when determining
whether to accept a new configuration or when calculating the cost
associated with moving from a current to a new configuration), energy
is associated with a configuration and is also stored in the {\it cfg}
class.

The class {\it MolecularStructureRosetta} provides the interface
between ROMEO and the Rosetta software suite.  This class, derived
from the {\it CfgForwardKinematics class} offers the ability to
extract a configuration from a Protein DataBank (PDB)~\citep{PDB03} file (which is the conventional way of storing information about a tertiary
structure) and to {\it score} a configuration using a selected Rosetta
scoring/energy functions.  Member functions are also provided to
perform forward kinematics on a {\it cfg} object, that is, placing the
protein in a particular configuration and obtaining the Cartesian
coordinates for each atom in a given representation.  ROMEO provides
support for Rosetta's centroid representation, which tracks heavy
backbone atoms and a side-chain centroid pseudo-atom.

\subsection*{Planners}

ROMEO utilizes sampling-based motion planning algorithms to explore
the configuration space of a protein.  Direct support is provided for
two popular sampling-based planners: the Rapidly Exploring Random Tree
(RRT)~\citep{RRT2001} and the Probabilistic RoadMap
(PRM)~\citep{KavrakiSvetskaLatombeOvermars96}.  As a proof of concept,
to illustrate ROMEO's plug and play architecture, the
FeLTR~\citep{ShehuOlsonIJRR10} and
SPRINT~\citep{MolloyShehuBMCStructBiol13} methods have also been
implemented in this release of ROMEO.  By expanding the sampler
classes, many variants of these planners can be implemented by other
developers.

These planners all share a common set of operations or core
components, such as: generating new configurations, determining the
validity of a new configuration, measuring the distance between two
configurations, and projecting configurations into the workspace.
ROMEO abstracts these operations using a set of base classes that
allow for easy ``plug and play'' replacements of these components.

\subsection*{Samplers and Offspring Generators}

ROMEO offers two classes, one for sampling and another for offspring
generation.  The {\it sampling} class is employed is during the
generation of at-random samples/configurations. Examples include
obtaining $q_{RAND}$ in the RRT planner and generating landmark
configurations to populate the roadmap/graph in PRM. The {\it
  offspring generator} class is used to \emph{modify} an existing
configuration, potentially by perturbing it in a given direction. This
is employed, for instance, when extending $q_{near}$ in the direction
of $q_{rand}$ in the RRT extend step, or when connecting landmark
configurations during the local planning step within the PRM
planner. ROMEO extends these baseline classes to support utilizing
Rosetta's molecular fragment libraries for structure and motion
computations.

\subsubsection*{Acceptors}

The {\it acceptor} class is used when deciding whether a new
configuration should be added to the graph or tree maintained by the
planner. A simple acceptor test would be to set a maximum energy value
for all configurations (thus, the acceptor verifies that new
configurations are below this threshold).  When studying molecular
transitions, the Metropolis criterion (which utilizes differences in
energy between two configurations) is commonly employed. For this
reason, ROMEO also provides an acceptor class based on MMC; in
particular, ROMEO implements the transition test utilized in the SPRINT~\citep{MolloyShehuBMCStructBiol13} and Transition-based RRT (T-RRT) planners~\citep{jailletcortes11, jaillet08}.

\subsubsection*{Distances}

Planners utilize a distance function to determine nearby
configurations. ROMEO comes with basic distance classes (Euclidean
distance), as well as distances commonly used in the study of proteins
(such as the least root-mean-square-distance, or lRMSD\citep{McLachlan72}.

\section*{Selected Examples of Applicability}

This section outlines two applications that utilize the ROMEO
framework.  The first example utilizes ROMEO to perform template-free
protein structure prediction by implementing a robotics-inspired
method known as FeLTR~\citep{ShehuOlsonIJRR10}. The second example
employs RRT to compute energetically-feasible paths that connect two
functionally-relevant configurations of the cyanovirin-n protein.  All
source code and scripts to run these examples are included with the
ROMEO distribution.

\subsection*{Structure Prediction}

This example showcases the FeLTR method~\citep{ShehuOlsonIJRR10} for
performing template-free protein structure prediction. We summarize how
ROMEO's ``plug and play'' architecture is utilized to easily implement
FeLTR.

\begin{figure}
\includegraphics[width=0.75\linewidth]{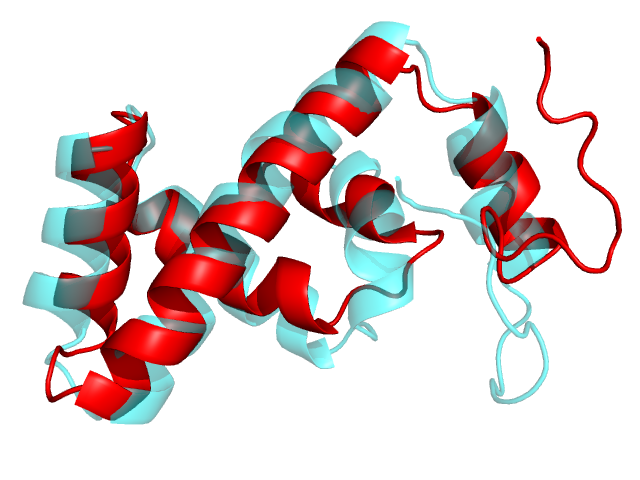}
\caption{The known native tertiary structure of the ibeta subdomain of
  the mu end DNA binding domain of phage mu transposase (found under
  PDB entry 2EZK) is drawn in transparent blue. The structure with the lowest lRMSD (of 3.3\AA) to this native structure (among all
  structures computed by FeLTR during a 2-hour execution) is also shown here, drawn in red.}
\label{fig:2ezkPrediction}
\end{figure}

\begin{figure}
\includegraphics[width=0.75\linewidth]{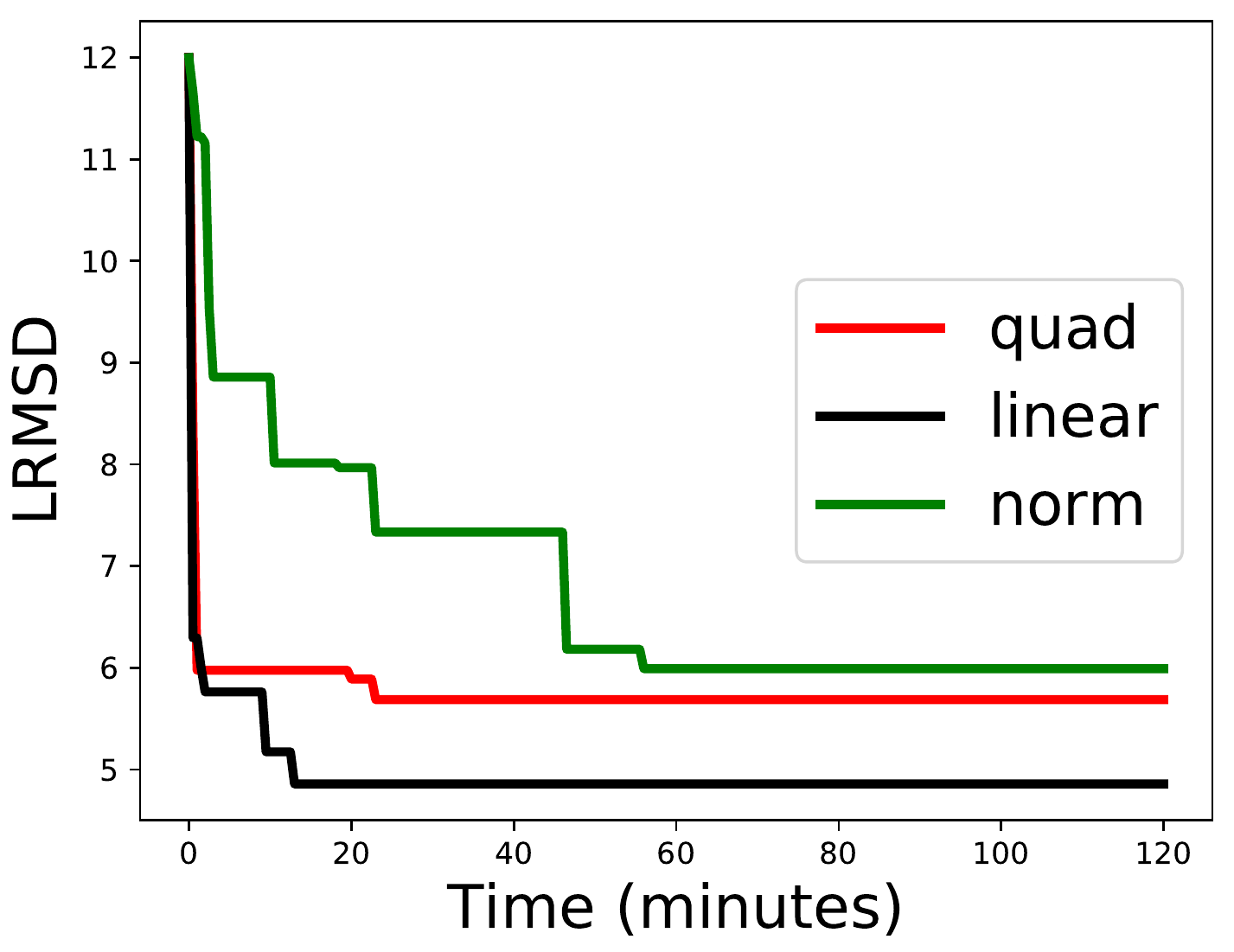}
\caption{The known native tertiary structure of the ibeta subdomain of
  the mu end DNA binding domain of phage mu transposase (found under
  PDB entry 2EZK) is drawn in transparent blue. The structure with the
  lowest lRMSD (of 3.3\AA) to this native structure (among all
  structures computed by FeLTR during a 2-hour execution) is also shown here, drawn in red.}
\label{fig:feltrRMSDProgress}
\end{figure}

As described in greater detail in~\citep{ShehuOlsonIJRR10}, FeLTr
employs an EST-like planner that expands the search tree iteratively
via selection and expansion operations and utilizes projection layers
for configuration selection. The {\it SelectionVertex} function of
ROMEO's {\it TreeSamplingBasedPlanner} class is overloaded to support
FeLTr's novel node selection technique that utilizes a low-dimensional
projection of a configuration. The {\it AddVertex} function is
overloaded to place new vertices into FeLTr's projection
layers. Changes to the planner are limited to the addition of $100$
lines of code (in addition to supporting code for computing projection
coordinates from configurations).  The ease with which this complex
method is implemented in ROMEO highlights the advantages of its
object-oriented design.

ROMEO's distribution provides the required scripts and configuration
files to run FeLTR to predict the structure of the ibeta subdomain of the mu
end DNA binding domain of phage mu transposase.  Figure~\ref{fig:2ezkPrediction} showcases the closest (in terms of lRMSD) sampled structure when compared to the native structure cataloged in the PDB.  Different weighting schemes can be explored when selecting nodes from FeLTR's low-dimensional projection, as described in \citep{MolloyShehuTCBB13}.  Figure~\ref{fig:feltrRMSDProgress} highlights some of the behavior of each of these weighting schemes.  The {\it quad} scheme utilizes a greedy strategy, selecting nodes with the lowest energies with high probability.  The {\it linear} and {\it norm} (Gaussian based) schemes approach the lower lRMSD structure more gradually, and have been shown for longer executions to provide closer samples to the native structure~\citep{MolloyShehuTCBB13}.

\subsection*{Motion Computation}

ROMEO can also be used to compute the motions that mediate
rearrangements between two distinct functionally-relevant
structures. We highlight this capability here on the cyanovirin-n
protein, where we treat two distinct structures (found under PDB IDs
2EZM and 1L5E) as start and goal structures located almost 16 \AA~lRMSD apart from one another.  We highlight here an implementation of the SPRINT
method~\citep{MolloyShehuBMCStructBiol13} with ROMEO.  

\begin{figure*}
\begin{tabular}{ccc}
\includegraphics[width=0.33\textwidth]{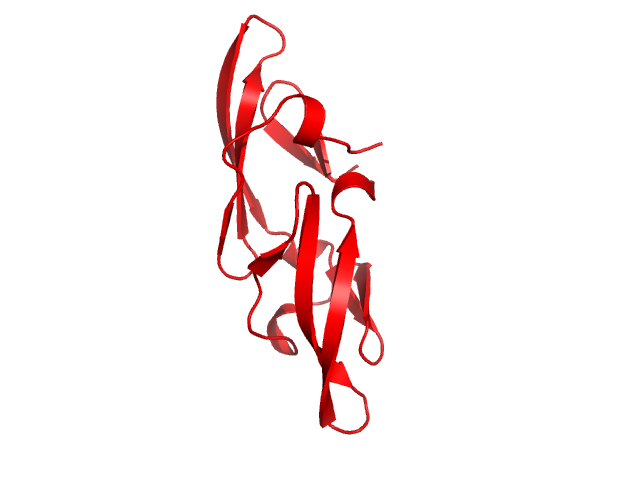} &
\includegraphics[width=0.33\textwidth]{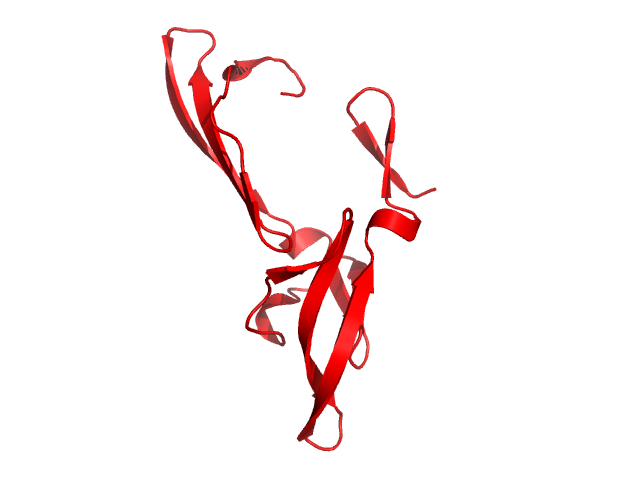} & 
\includegraphics[width=0.33\textwidth]{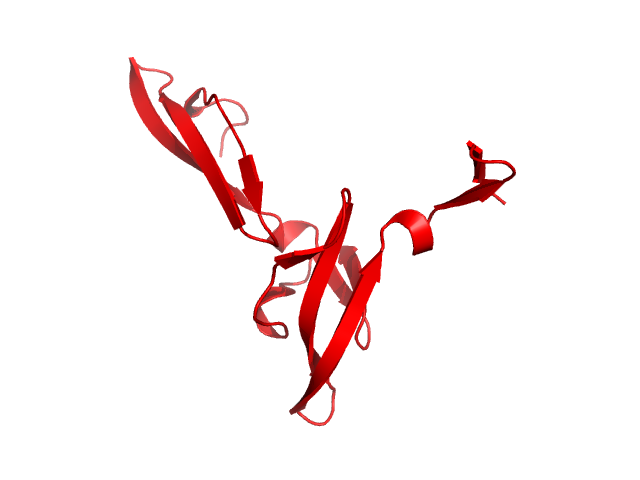} \\

\includegraphics[width=0.33\textwidth]{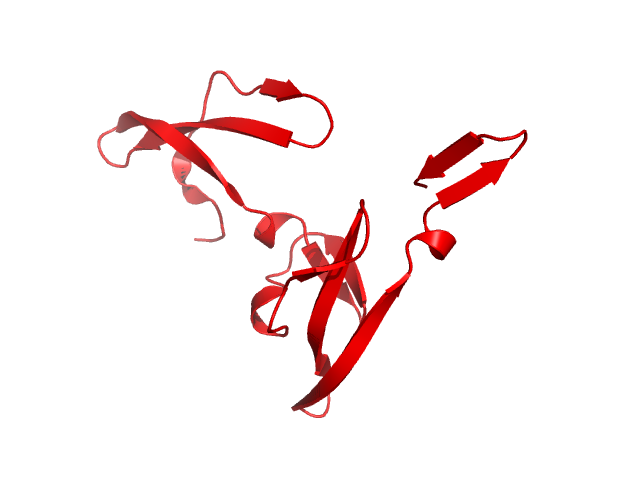} &
\includegraphics[width=0.33\textwidth]{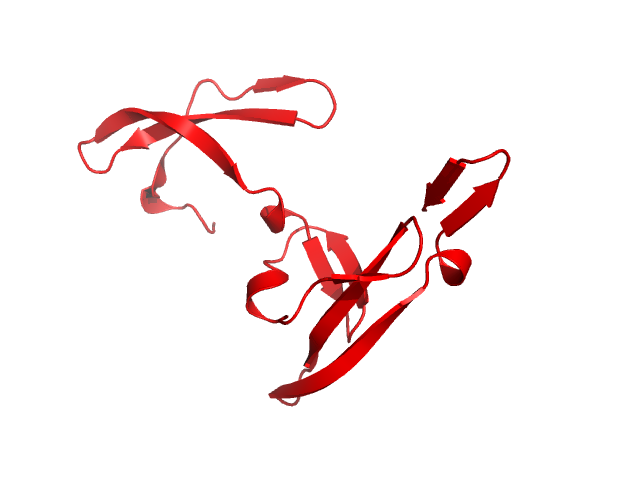} &
\includegraphics[width=0.33\textwidth]{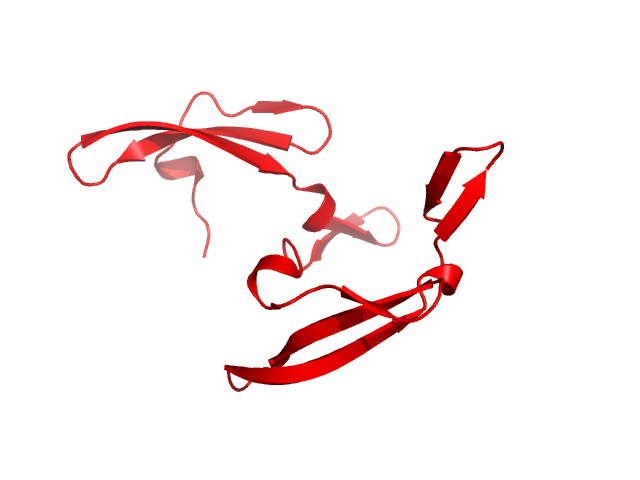} \\
\end{tabular}
\caption{A few ROMEO-computed structures are shown along a path
that illustrates the rearrangement of the cyanovirin-n protein between two distinct, known structures (with PDB IDs 2EZM and 1L5E).  The computed
path consists of $89$ intermediate structures, of which 6 are shown here. The first stucture (top left) is the one under PDB ID 2EZM. The last structure is that under PDB ID 1L5E. The other four structures are selected from the ROMEO-computed path to illustrate the rearrangement of the protein between the given start and goal structures.}
\label{fig:CVNPathPics}
\end{figure*}

We executed ROMEO for 12 hours and tested two different energy acceptors.  Rosetta's score3 energy scheme was utilized with the radius of gyration terms disabled (since this rewards more compact structures).  The first acceptor limited the acceptable energy of a configuration to no higher than 60 Rosetta Energy Units (REUs).  The second scheme utilized the Metropolis criterion, setting the temperature, such that an increase of $10$ REUs had a 0.1 probability of being accepted. Each scheme resulted in pathways that ended with configurations (structures) within 3 \AA~lRMSD of the goal configuration.  The energy profiles for each are shown in Figure ref{fig:CVNPathEnergy}.
As expected, the Metropolis acceptor (MMC) shows a more gradually-increasing and overall a lower-energy path compared to the max energy acceptor.  A few sample structures along the pathway computed from the MMC execution are showcased in Figure~\ref{fig:CVNPathPics}.

\begin{figure}
\includegraphics[width=0.8\linewidth]{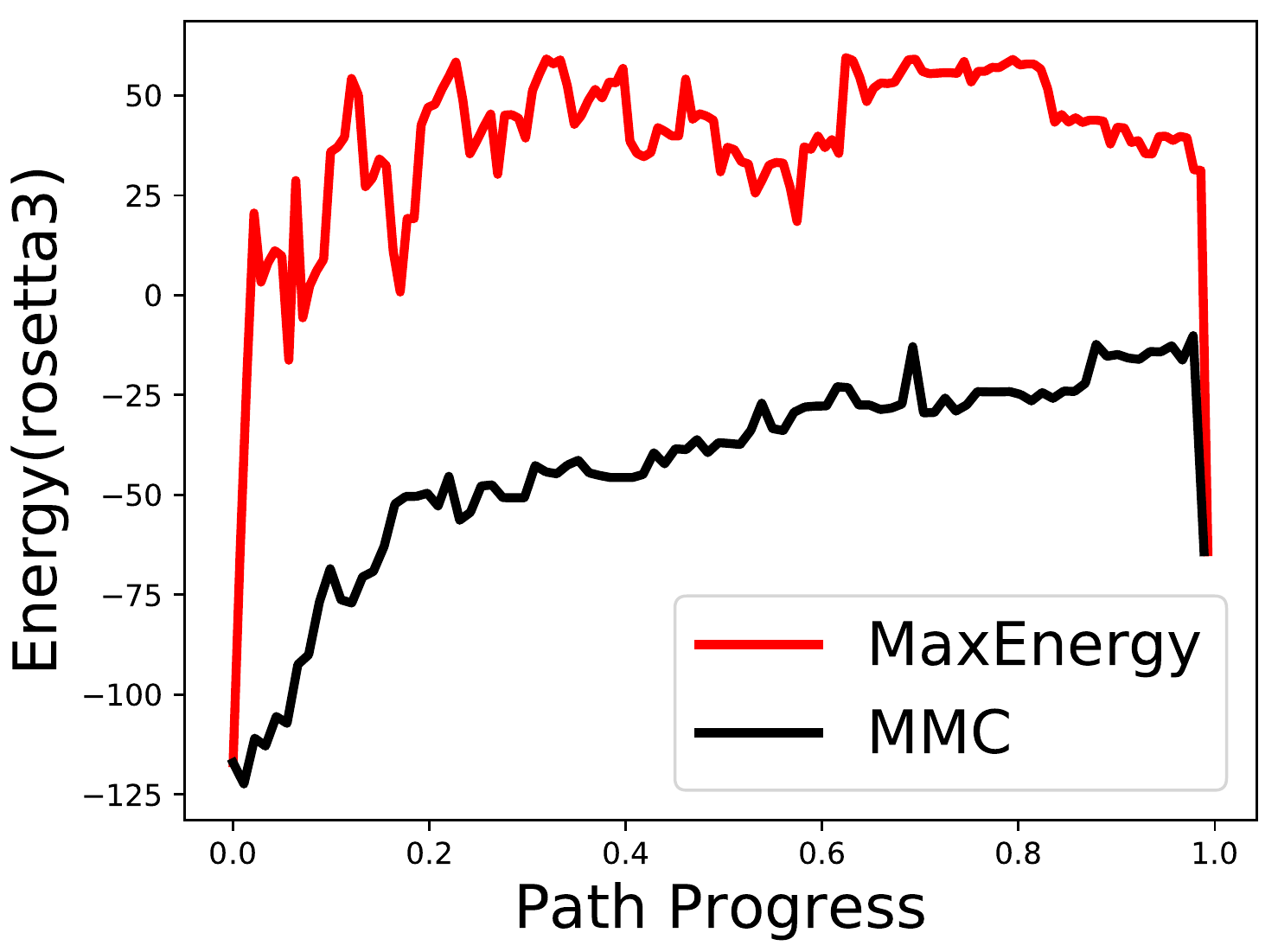}
\caption{Two energy acceptance strategies are utilized, MMC (Metropolis Monte Carlo) and a scheme that sets the maximum energy of a structure during
ROMEO's exploration of paths connecting two distinct, known structures of the cyanovirin-n protein.  The MMC scheme yields a more energetically-feasible pathway that gradually rises in energy compared to the maximum energy threshold strategy.}
\label{fig:CVNPathEnergy}
\end{figure}

\end{document}